\documentstyle[aps]{revtex}

\draft

\begin{document}
\title{Geometry effects in confined space}
\author{Wu-Sheng Dai \thanks{
daiwusheng@tju.edu.cn}}
\address{Department of Physics, Tianjin University, Tianjin 300072, P. R.
China\\
LuiHui Center for Applied Mathematics, Nankai University \&
Tianjin University, Tianjin 300072, P. R. China}
\author{Mi Xie \thanks{
xiemi@mail.tjnu.edu.cn}}
\address{Department of Physics, Tianjin Normal University, Tianjin 300074,
P. R. China}
\date{}
\maketitle

\begin{abstract}
In this paper we calculate some exact solutions of the grand partition
functions for quantum gases in confined space, such as ideal gases in two-
and three-dimensional boxes, in tubes, in annular containers, on the lateral
surface of cylinders, and photon gases in three-dimensional boxes. Based on
these exact solutions, which, of course, contain the complete information
about the system, we discuss the geometry effect which is neglected in the
calculation with the thermodynamic limit $V\rightarrow \infty $, and analyze
the validity of the quantum statistical method which can be used to
calculate the geometry effect on ideal quantum gases confined in
two-dimensional irregular containers. We also calculate the grand partition
function for phonon gases in confined space. Finally, we discuss the
geometry effects in realistic systems.
\end{abstract}


\section*{I. Introduction}

In statistical mechanics, for seeking the sum over all possible states, we
always take the approximation that the volume of the system tends to
infinity. In so doing we have lost all information about the geometry
property of the system because in such an approximation the spectrum of
single-particle states is continuous while the total number of states is
independent of the shape of the boundary and simply proportional to the
volume of the system. However, the properties of some systems found recently
are shape dependent and sensitive to the topology \cite%
{Potempa,Braun,Kravtsov}. In Ref. \cite{Ours}, we developed a method for
calculating the effects of boundary and topology on ideal Bose and Fermi
gases in confined space.

In the thermodynamic limit, one can replace the summation over states by an
integral: $\sum_{s}\longrightarrow \int d\varepsilon \rho (\varepsilon )$,
where

\begin{equation}
\rho (\varepsilon )=V\frac{(2\pi m)^{d/2}}{h^{d}\Gamma (d/2)}\varepsilon
^{d/2-1}  \label{e1.5}
\end{equation}%
is the density of states of a $d$-dimensional gas of dispersion $%
E=p^{2}/(2m) $. However, if the system is enclosed in a finite volume, the
structure of the phase space will be changed. In confined space, the
spectrum of single-particle states will depend on the shape of the boundary.
This is important if the thermal wavelength of particles is comparable to
the size of the container, that is to say the replacement of the summation
with the integral with the density of states Eq.(\ref{e1.5}) is a good
approximation only when the volume is large enough so that the particle can
not feel the boundary.

For taking into account geometry effects, in the preceding paper \cite{Ours}
we proposed that for the case of a two-dimensional ideal quantum gas
confined in an irregular container, the density of states, after neglecting
the topological contribution, can be written as

\begin{equation}
\rho (\varepsilon )=S\frac{2\pi m}{h^{2}}-\frac{1}{4}L\frac{(2m)^{1/2}}{h}%
\frac{1}{\varepsilon ^{1/2}},  \label{e1.6}
\end{equation}%
where $S$ is the area and $L$ the perimeter of the container. Moreover, for
the case of a three-dimensional ideal quantum gas in a long tube, of which
all transverse cross-sections keep the same, the density of states is
\begin{equation}
\rho (\varepsilon )=L_{z}S\frac{(2\pi m)^{3/2}}{h^{3}\Gamma (3/2)}%
\varepsilon ^{1/2}-\frac{1}{4}L_{z}L\frac{2\pi m}{h^{2}}.  \label{e1.8}
\end{equation}%
$S$ and $L$ here denote the area and perimeter of the transverse
cross-section of the tube, respectively, and the length of the tube $L_{z}$
is made sufficiently large so that the $z$-component of the momentum $p_{z}$
can be considered to be continuous. The second terms of Eqs. (\ref{e1.6})
and (\ref{e1.8}) describe the contributions from the boundary.

In the present paper, for analyzing geometry effects, we will calculate some
exact solutions for ideal gases and photon gases in various kinds of
confined space, and provide an approximate result for phonon gases in
confined space. By comparing these exact and approximate solutions with the
result obtained in the thermodynamic limit $V\rightarrow \infty $, we can
extract the information about boundary shape and topology. Furthermore, the
validity of the method presented in Ref. \cite{Ours} can be further
justified by comparing it with these exact solutions.

It can be found that the corrections to the standard results which are on
the order $\lambda /L$ where $\lambda $ is the thermal wavelength and $L$
denotes the linear size of the system. At the end of this paper, we will
discuss such geometry effects in realistic systems.

In Sec. II, we discuss the method which will be used in this paper for
calculating geometry effects. In Sec. III, we calculate the exact solutions
for ideal gases in two-dimensional boxes, three-dimensional tubes and boxes.
By comparing Eqs. (\ref{e1.6}) and (\ref{e1.8}) with these exact solutions,
we can see that the densities of states given by Eqs. (\ref{e1.6}) and (\ref%
{e1.8}) are very good approximations for discussing ideal gases in confined
space. Moreover, in this section, we also consider the geometry effect on
the Fermi energy. In Sec. IV, we calculate the exact solutions for ideal
gases in annular containers. The result shows no boundary effects in such
systems even the scales of the systems are very small. In Sec. V, we
calculate the exact solutions for photon gases in confined space, and
compare the thermodynamic quantities with the standard result of black-body
radiation. In Sec. VI, we calculate an approximate result for phonon gases
in confined space. Based on this result, we can analyze the boundary effect
on the lattice specific heat. In Sec. VII, we analyze the geometry effects
in realistic systems and compare the influence of geometry effects with the
influence of fluctuations and interparticle interactions. The conclusions
are summarized in Sec. VIII while some expressions of thermodynamic
quantities are given in Appendix A.

\section*{II. The method for calculating geometry effects}

For an exact study of statistical mechanics in confined space, the geometry
effect has to be reckoned in. To seek the exact solution of a grand
potential, we need to find a method to perform the summation over all
possible states because the energy spectrum is discrete in confined space.
In the following we will solve the sum by using the Euler-MacLaurin formula
\cite{Abramowitz}:

\begin{equation}
\sum\limits_{n=0}^{\infty }F(n)=\int_{0}^{\infty }F(n)dn+\frac{1}{2}F(0)-%
\frac{1}{2!}B_{2}F^{\prime }(0)-\frac{1}{4!}B_{4}F^{\prime \prime \prime
}(0)+\cdots ,  \label{e2.5}
\end{equation}%
where $B_{\nu }$ are Bernoulli Numbers: $B_{2}=1/6$, $B_{4}=-1/30$, $\cdots $%
.

The Euler-MacLaurin formula converts a sum to a series. Generally speaking,
the Euler-MacLaurin formula can be used as an approximate method to solve
sums approximately. However, in some special cases, this series has only a
finite number of non-zero terms, i.e., $F^{(2\nu +1)}(0)=0$ when $\nu $ is
greater than a certain value (note that $B_{2\nu +1}=0$ when $\nu >0$), and
one can obtain the exact solution of the sum.

Moreover, a more general expression for the Euler-MacLaurin formula is

\begin{equation}
\sum\limits_{n=k}^{m-1}F(n)=\int_{k}^{m}F(n)dn-\frac{1}{2}\left[ F(m)-F(k)%
\right] +\frac{B_{2}}{2!}\left[ F^{\prime }(m)-F^{\prime }(k)\right] +\frac{%
B_{4}}{4!}\left[ F^{\prime \prime \prime }(m)-F^{\prime \prime \prime }(k)%
\right] +\cdots ,  \label{e2.6}
\end{equation}%
which will be used to discuss the phonon gas in confined space.

In statistical mechanics, one always replaces the summation over states by
an integral in the thermodynamic limit $V\rightarrow \infty $. This
treatment is equivalent to replacing the summation only by the integral on
the right-hand side of Eq. (\ref{e2.5}) but neglecting the contribution from
the rest terms. The information about the geometry of the system is
contained in the summation over states since the structure of the spectrum
of noninteracting particles is determined by the system geometry. The
geometry information, however, will be lost after the replacement of the
summation with the integral; in other words, the integral in Eq. (\ref{e2.5}%
) does not contain geometry information, that is to say, the rest terms in
Eq. (\ref{e2.5}) correspond to geometry effects.

It is natural to expect that if there is no boundary, there is no boundary
effect at all. No boundary implies periodic boundary conditions; in other
words, there will be no boundary effect if we apply periodic boundary
conditions. For example, we will show that there is no boundary effect in
the system of an ideal gas in an annular container.

The prerequisite for applying the Euler-MacLaurin formula to solving the sum
over states is that the energy spectrum of the system must be known. Whether
one can obtain an exact solution or not depends on whether the series has a
finite number of non-zero terms. If there are infinite non-zero $F^{(2\nu
+1)}(k)$, one obtains only an approximate solution. In the following we will
calculate some exact and approximate solutions for ideal gases, photon gases
and phonon gases in confined space.

\section*{III. Exact solutions for ideal gases in confined space}

In this section, we first calculate some exact solutions for ideal gases in
two- and three-dimensional confined space. Based on these exact solutions we
then discuss the validity of the approximate method provided in Ref. \cite%
{Ours}, which can be used to calculate the geometry effect in the system
with an irregular boundary. Moreover, as an application, we discuss the
boundary effect on the Fermi energy.

\subsection*{1. Exact solutions for ideal gases in two-dimensional cases}

In this part, by using the Euler-MacLaurin formula, we calculate the
boundary effects on ideal gases in two two-dimensional cases exactly: a)
ideal gases in two-dimensional boxes, and b) ideal gases on the lateral
surface of cylinders.

\subsubsection*{a) Exact solutions for ideal gases in two-dimensional boxes}

Next we discuss the exact solution for ideal gases in two-dimensional boxes
by using the Euler-MacLaurin formula.

The energy spectrum of a particle in a two-dimensional rectangular box of
sides $L_{x}$ and $L_{y}$ is

\begin{equation}
E(n_{x},n_{y})=\frac{\pi ^{2}\hbar ^{2}}{2m}\left( \frac{n_{x}^{2}}{L_{x}^{2}%
}+\frac{n_{y}^{2}}{L_{y}^{2}}\right) ,~~~~n_{x},n_{y}=1,2\cdots .
\end{equation}%
Then the grand potential of the system can be expressed as

\begin{equation}
\ln \Xi =\mp \sum\limits_{n_{x}=1}^{\infty }\sum\limits_{n_{y}=1}^{\infty
}\ln \left[ 1\mp ze^{-\beta E(n_{x},n_{y})}\right] .
\end{equation}%
In this equation and following, the upper sign stands for bosons and the
lower sign for fermions. By use of the Euler-MacLaurin formula, we can
perform the summation exactly:

\begin{eqnarray}
\ln \Xi &=&\mp \int\limits_{0}^{\infty }\int\limits_{0}^{\infty
}dn_{x}dn_{y}\ln \left[ 1\mp ze^{-\beta E(n_{x},n_{y})}\right] \pm \frac{1}{2%
}\int\limits_{0}^{\infty }dn_{x}\ln \left[ 1\mp ze^{-\beta E(n_{x},0)}\right]
\nonumber \\
&&\pm \frac{1}{2}\int\limits_{0}^{\infty }dn_{y}\ln \left[ 1\mp ze^{-\beta
E(0,n_{y})}\right] \mp \frac{1}{4}\ln (1\mp z).
\end{eqnarray}%
We have

\begin{equation}
\ln \Xi =\frac{S}{\lambda ^{2}}h_{2}(z)-\frac{1}{4}\frac{L}{\lambda }%
h_{3/2}(z)+\frac{1}{4}h_{1}(z),  \label{e2.8}
\end{equation}%
where $S=L_{x}L_{y}$ is the area and $L=2(L_{x}+L_{y})$ the perimeter of the
box, $\lambda =h/\sqrt{2\pi mkT}$ is the thermal wavelength, and the function

\begin{equation}
h_{\sigma }(z)=\frac{1}{\Gamma (\sigma )}\int_{0}^{\infty }\frac{x^{\sigma
-1}}{z^{-1}e^{x}\mp 1}dx
\end{equation}%
equals the\ Bose-Einstein integral $g_{\sigma }(z)$ or the Fermi-Dirac
integral $f_{\sigma }(z)$ in boson or fermion case, respectively.

Note that Eq. (\ref{e2.8}) is the exact solution for the grand potential.
The first term is just the result obtained in the thermodynamic limit $%
S\rightarrow \infty $; the boundary effect is described by the second term
which is proportional to the perimeter $L$. Comparing Eq. (\ref{e2.8}) with
the corresponding approximate result given in Ref. \cite{Ours}

\begin{equation}
\ln \Xi =\frac{S}{\lambda ^{2}}h_{2}(z)-\frac{1}{4}\frac{L}{\lambda }%
h_{3/2}(z)+\frac{\chi }{6}h_{1}(z),  \label{e2.10}
\end{equation}%
we can see that the first two terms are completely equal to each other and
the error introduced by the approximation appears only in the third term
which describes the topology effect. The contribution from the topology
effect is proportional to the Euler-Poincar\'{e} characteristic number $\chi
$, in this case $\chi =1-r=1$ since $r$, the number of holes in this
two-dimensional box, equals zero. From this, we can see the close analogy
between the two results expressed in Eqs. (\ref{e2.8}) and (\ref{e2.10}).
This close correspondence, notice that Eq. (\ref{e2.8}) is an exact
solution, can be regarded as an evidence of the validity of the method given
in Ref. \cite{Ours}.

\subsubsection*{b) Exact solutions for ideal gases on the lateral surface of
cylinders}

Consider an ideal gas confined on the lateral surface of a cylinder with
radius $R$ and length $L$. Such a case is just the same as the ideal gas
confined in a two-dimensional box of sides $L_{x}=2\pi R$ and $L_{y}=L$, but
the wavevector along $x$ is determined by periodic boundary conditions:

\begin{equation}
k_{x}=\frac{n_{x}}{R},~~~n_{x}=0,\pm 1,\pm 2,\cdots ,
\end{equation}%
while the wavevector\ along $y$ is determined by fixed-end boundary
conditions:

\begin{equation}
k_{y}=\frac{n_{y}\pi }{L},~~~n_{y}=1,2,\cdots .
\end{equation}%
The energy spectrum of a particle is

\begin{equation}
E(n_{x},n_{y})=\frac{\hbar ^{2}}{2m}\left( k_{x}^{2}+k_{y}^{2}\right) .
\end{equation}%
Then the grand potential can be calculated exactly by use of the
Euler-MacLaurin formula:

\begin{eqnarray}
\ln \Xi &=&\mp \sum\limits_{n_{x}}\sum\limits_{n_{y}}\ln \left[ 1\mp
ze^{-\beta E(n_{x},n_{y})}\right]  \nonumber \\
&=&\frac{S}{\lambda ^{2}}h_{2}(z)-\frac{1}{4}\frac{L}{\lambda }h_{3/2}(z),
\end{eqnarray}%
where $S=2\pi RL$ is the area of the surface, and $L=2\cdot 2\pi R=4\pi R$
is the total length of the sides.

Furthermore, we can calculate the corresponding thermodynamic quantities
directly, e.g., the specific heat

\begin{equation}
\frac{C_{V}}{Nk}=\sigma \,\left[ 2\frac{h_{2}(z)}{h_{1}(z)}-\gamma \frac{%
h_{1}(z)}{h_{0}(z)}\right] -\frac{1}{\sqrt{N}}\frac{L}{\sqrt{S}}\sqrt{\sigma
}\left[ \frac{3}{16}\frac{h_{3/2}(z)}{h_{1}^{1/2}(z)}-\frac{1}{8}\gamma
\frac{h_{1}^{1/2}(z)h_{1/2}(z)}{h_{0}(z)}\right] ,  \label{e2.18}
\end{equation}%
where
\[
\sigma =\left[ \sqrt{1+\frac{1}{64N}\frac{L^{2}}{S}\,\frac{h_{1/2}^{2}(z)}{%
h_{1}(z)}}-\frac{1}{8\sqrt{N}}\frac{L}{\sqrt{S}}\frac{h_{1/2}(z)}{%
h_{1}^{1/2}(z)}\right] ^{-2},
\]%
\[
\gamma =\frac{1-\frac{1}{8\sqrt{N}}\frac{L}{\sqrt{S}}\frac{h_{1/2}(z)}{%
h_{1}^{1/2}(z)}\frac{1}{\sqrt{\sigma }}}{1-\frac{1}{4\sqrt{N}}\frac{L}{\sqrt{%
S}}\frac{h_{1}^{1/2}(z)h_{-1/2}(z)}{h_{0}(z)}\frac{1}{\sqrt{\sigma }}}.
\]%
The following relation is used in the calculation:
\begin{equation}
\frac{\partial z}{\partial T}=-\frac{z}{T}\frac{h_{1}(z)}{h_{0}(z)}\gamma .
\end{equation}

\subsection*{2. Exact solutions for ideal gases in three-dimensional cases}

We next calculate the boundary effects on ideal gases in two
three-dimensional cases exactly: a) ideal gases in three-dimensional tubes
with rectangular transverse cross-sections, and b) ideal gases in
three-dimensional boxes.

\subsubsection*{a) Ideal gases in three-dimensional tubes}

Consider an ideal gas enclosed in a three-dimensional tube with a
rectangular transverse cross-section of sides $L_{x}$ and $L_{y}$. The $z$%
-component of the momentum $p_{z}$ is continuous since the length of the
tube $L_{z}$ is made sufficiently large. In this case, the energy spectrum
of a particle in the tube can be written as

\begin{equation}
E(n_{x},n_{y},p_{z})=\frac{\pi ^{2}\hbar ^{2}}{2m}\left( \frac{n_{x}^{2}}{
L_{x}^{2}}+\frac{n_{y}^{2}}{L_{y}^{2}}\right) +\frac{p_{z}^{2}}{2m}
,~~~~~~~n_{x},n_{y}=1,2\cdots .
\end{equation}
The grand potential is

\begin{equation}
\ln \Xi =\mp \int_{-\infty }^{\infty }\frac{L_{z}dp_{z}}{h}
\sum\limits_{n_{x}=1}^{\infty }\sum\limits_{n_{y}=1}^{\infty }\ln \left[
1\mp ze^{-\beta E(n_{x},n_{y},p_{z})}\right] .
\end{equation}
Here, we have converted the summation over $p_{z}$ into an integral since $%
p_{z}$ is continuous. Using the Euler-MacLaurin formula, we can perform the
summations over $n_{x}$ and $n_{y}$ exactly:

\begin{equation}
\ln \Xi =\frac{V}{\lambda ^{3}}h_{5/2}(z)-\frac{1}{4}\frac{S}{\lambda ^{2}}%
h_{2}(z)+\frac{1}{4}\frac{L_{z}}{\lambda }h_{3/2}(z).  \label{e3.8}
\end{equation}%
Here the volume $V=L_{x}L_{y}L_{z}$ and the area of the lateral surface of
the tube $S=L_{z}L$ where $L=2(L_{x}+L_{y})$ is the perimeter of the
transverse cross-section.

The first two terms of the exact solution Eq. (\ref{e3.8}) are consistent
with the approximate result given in Ref. \cite{Ours}:

\begin{equation}
\ln \Xi =\frac{V}{\lambda ^{3}}h_{5/2}(z)-\frac{1}{4}\frac{S}{\lambda ^{2}}%
h_{2}(z)+\frac{\chi }{6}\frac{L_{z}}{\lambda }h_{3/2}(z),  \label{e3.10}
\end{equation}%
where $\chi $ is the Euler-Poincar\'{e} characteristic number of the
transverse cross-section of the tube and in this case $\chi =1$. Comparing
Eq. (\ref{e3.8}) with Eq. (\ref{e3.10}) we can see that, just as the
two-dimensional case, the error only appears in the term corresponding to
the topology effect.

\subsubsection*{b) Ideal gases in three-dimensional boxes}

The energy spectrum of a particle in a three-dimensional box of sides $L_{x}$%
, $L_{y}$ and $L_{z}$ is

\begin{equation}
E(n_{x},n_{y},n_{z})=\frac{\pi ^{2}\hbar ^{2}}{2m}\left( \frac{n_{x}^{2}}{%
L_{x}^{2}}+\frac{n_{y}^{2}}{L_{y}^{2}}+\frac{n_{z}^{2}}{L_{z}^{2}}\right)
,~~~~~~~n_{x},n_{y},n_{z}=1,2\cdots .
\end{equation}%
The grand potential of an ideal gas in this box is

\begin{equation}
\ln \Xi =\mp \sum\limits_{n_{x}=1}^{\infty }\sum\limits_{n_{y}=1}^{\infty
}\sum\limits_{n_{z}=1}^{\infty }\ln \left[ 1\mp ze^{-\beta
E(n_{x},n_{y},n_{z})}\right] .
\end{equation}%
By use of the Euler-MacLaurin formula, we can calculate the sum exactly:

\begin{eqnarray}
\ln \Xi &=&\mp \int\limits_{0}^{\infty }\int\limits_{0}^{\infty
}\int\limits_{0}^{\infty }dn_{x}dn_{y}dn_{z}\ln \left[ 1\mp ze^{-\beta
E(n_{x},n_{y},n_{z})}\right] \pm \frac{1}{2}\int\limits_{0}^{\infty
}\int\limits_{0}^{\infty }dn_{y}dn_{z}\ln \left[ 1\mp ze^{-\beta
E(0,n_{y},n_{z})}\right]  \nonumber \\
&&\pm \frac{1}{2}\int\limits_{0}^{\infty }\int\limits_{0}^{\infty
}dn_{x}dn_{z}\ln \left[ 1\mp ze^{-\beta E(n_{x},0,n_{z})}\right] \pm \frac{1%
}{2}\int\limits_{0}^{\infty }\int\limits_{0}^{\infty }dn_{x}dn_{y}\ln \left[
1\mp ze^{-\beta E(n_{x},n_{y},0)}\right]  \nonumber \\
&&\mp \frac{1}{4}\int\limits_{0}^{\infty }dn_{x}\ln \left[ 1\mp ze^{-\beta
E(n_{x},0,0)}\right] \mp \frac{1}{4}\int\limits_{0}^{\infty }dn_{y}\ln \left[
1\mp ze^{-\beta E(0,n_{y},0)}\right]  \nonumber \\
&&\mp \frac{1}{4}\int\limits_{0}^{\infty }dn_{z}\ln \left[ 1\mp ze^{-\beta
E(0,0,n_{z})}\right] \pm \frac{1}{8}\ln (1\mp z).
\end{eqnarray}%
Then we have

\begin{equation}
\ln \Xi =\frac{V}{\lambda ^{3}}h_{5/2}(z)-\frac{1}{4}\frac{S}{\lambda ^{2}}%
h_{2}(z)+\frac{1}{16}\frac{L}{\lambda }h_{3/2}(z)-\frac{1}{8}h_{1}(z),
\label{e3.18}
\end{equation}%
where $V=L_{x}L_{y}L_{z}$ is the volume, $%
S=2(L_{x}L_{y}+L_{y}L_{z}+L_{z}L_{x})$ the area of the surface and $L=$ $%
4\left( L_{x}+L_{y}+L_{z}\right) $ the total length of the sides of the box.

We know that a box will change to a tube if its one side, e.g., $L_{z}$, is
made very long. Assuming $L_{z}\gg L_{x},L_{y}$, we then can take the
approximation $S\simeq 2(L_{y}+L_{x})L_{z}$ and $L\simeq 4L_{z}$. It is easy
to see that in this case Eq. (\ref{e3.18}) is the same as Eq. (\ref{e3.8})
except the last term which is negligibly small. In above discussions, we
only concentrate on the case that the temperature of the system is higher
than the critical temperature.

\subsection*{3. The validity of the approximate method for calculating
geometry effects with irregular boundaries}

In Ref. \cite{Ours}, for calculating geometry effects, we first expand the
grand potential $\ln \Xi =\mp \sum\limits_{s}\ln (1\mp ze^{-\beta \epsilon
_{s}})$ as a series of $ze^{-\beta \epsilon _{s}}$, and then sum over all
terms of the expansion after calculating these terms by use of the result
given by Kac \cite{Kac}. Clearly, the validity of this treatment depends on
the value of the expansion parameter $ze^{-\beta \epsilon _{s}}$. Strictly
speaking, however, when the fugacity $z>1$, such as in Fermi-Dirac
statistics, the grand potential $\ln \Xi $ can not be expanded in such a
way. In the preceding paper \cite{Ours} we perform the summation of the
expansion of the grand potential without proving the convergence of the
series, but, instead, we analyze the validity of the result by comparing it
with the grand potential of an ideal quantum gas in free space and show that
the grand potential in free space, whose validity is fully accepted, is just
the zeroth-order approximation of our result. With the help of the exact
solutions obtained above, we can make further improvement on the analysis of
the validity of this treatment.

Comparing the results obtained by the approximate method developed in Ref.
\cite{Ours} with the exact solutions calculated in this section, we can see
that this approximate method is valid for calculating the geometry effects
in two- and three-dimensional ideal gas systems. So far we have applied two
methods to performing the summation in the grand potential for ideal gases
in confined space: One is an approximate method \cite{Ours} based on a
mathematical work given by Kac \cite{Kac}. Another is an exact one based on
the Euler-MacLaurin formula. The approximate method applies equally well to
the exact method in calculating the boundary effect, and the error
introduced by the approximation appears only in the topology terms and is
indeed small.

\subsection*{4. The boundary effect on the Fermi energy}

Using the result obtained above, we can consider the boundary effect on the
Fermi energy. From Eq. (\ref{e3.18}), we can obtain

\begin{equation}
N=g\left[ \frac{V}{\lambda ^{3}}f_{3/2}(z)-\frac{1}{4}\frac{S}{\lambda ^{2}}%
f_{1}(z)+\frac{1}{16}\frac{L}{\lambda }f_{1/2}(z)-\frac{1}{8}\frac{z}{z+1}%
\right] ,  \label{e3.28}
\end{equation}%
where $g$ is a weight factor that arises from the internal structure of the
particles (the number of internal degrees of freedom).

The Fermi energy $\mu _{0}$ is the energy of the topmost filled level in the
ground state of the $N$ electron system, so

\begin{equation}
N=g\left[ V\frac{1}{6\pi ^{2}}\left( \frac{\hbar ^{2}}{2m}\right) ^{-3/2}\mu
_{0}^{3/2}-\frac{1}{4}S\frac{1}{4\pi }\left( \frac{\hbar ^{2}}{2m}\right)
^{-1}\mu _{0}+\frac{1}{16}L\frac{1}{\pi }\left( \frac{\hbar ^{2}}{2m}\right)
^{-1/2}\mu _{0}^{1/2}-\frac{1}{8}\right] .
\end{equation}%
The Fermi energy can be exactly calculated from this equation; however, for
clarity we only calculate the next-to-leading-order correction:

\begin{eqnarray}
\mu _{0} &\simeq &\frac{\hbar ^{2}}{2m}\left( \frac{2}{g}3\pi ^{2}\frac{N}{V}%
\right) ^{2/3}+\frac{\hbar ^{2}}{2m}\frac{1}{2}\frac{S}{V}\left( \frac{3\pi
^{5}}{4g}\frac{N}{V}\right) ^{1/3}  \label{e3.38} \\
&=&\mu _{0}^{free}+\mu _{0}^{boundary}.
\end{eqnarray}%
Note that, for an electron gas the spin weight is $g=2$. Here $\mu
_{0}^{free}$ is just the Fermi energy in free space, and $\mu _{0}^{boundary}
$ reflects the influence of the boundary. It is easy to see that the
existence of a boundary enhances the Fermi energy, and the Fermi energy
increases with the decreasing size of the system. This result is not
difficult to understand: In a finite size system, ground state energy and
energy level spacing grow larger as the size of the system decreases.

\section*{IV. Exact solutions for ideal gases in annular containers}

The exact solutions for ideal gases in annular containers can also be
calculated with the help of the Euler-MacLaurin formula. In this section, we
consider an ideal quantum gas confined in an one-dimensional ring of radius $%
R$. The energy of the particle in such an annular container is determined by
periodic boundary conditions:%
\begin{equation}
\varepsilon _{n}=\frac{\hbar ^{2}}{2mR^{2}}n^{2},~~~n=0,\pm 1,\pm 2,\cdots .
\end{equation}%
The grand potential of the ideal gas is

\begin{eqnarray}
\ln \Xi &=&\mp \sum\limits_{n=-\infty }^{\infty }\ln \left( 1\mp ze^{-\beta
\varepsilon _{n}}\right)  \nonumber \\
&=&\mp \ln \left( 1\mp z\right) \mp 2\sum\limits_{n=1}^{\infty }\ln \left[
1\mp ze^{-\beta n^{2}\hbar ^{2}/\left( 2mR^{2}\right) }\right] .
\label{e8.8}
\end{eqnarray}%
The summation over the discrete parameter $n$ can be converted to an
integral exactly, leading to

\begin{eqnarray}
\ln \Xi &=&\mp 2\int\nolimits_{0}^{\infty }dn\ln \left[ 1\mp ze^{-\beta
n^{2}\hbar ^{2}/\left( 2mR^{2}\right) }\right]  \nonumber \\
&=&\frac{L}{\lambda }h_{3/2}(z),
\end{eqnarray}%
where $L=2\pi R$ is the perimeter of the container. There is no boundary
effect because, for a ring, there is no boundary. This is an interest
result. Generally speaking, for a finite perimeter $L$, the energy levels $%
\varepsilon _{n}$ are always discrete, so the summation in Eq. (\ref{e8.8})
has to be performed approximately by replacing it by an integral: $%
\sum\limits_{n=-\infty }^{\infty }\rightarrow \int_{-\infty }^{\infty }dn$.
This replacement is based on the assumption that the momentum is continuous,
which is valid only when $L\rightarrow \infty $. In many cases, such a
treatment is an approximation that neglects the influence of the boundary;
however, in an annular container this treatment gives an exact solution.

This result implies that, in a ring, e.g., an ideal electron gas in a
conductor ring, even the perimeter is very small, then the energy level
spacings are very large, we can replace the summation by an integral safely.
The thermodynamic behavior of an ideal gas in a ring is the same as that in
infinite free space. Generally, it is easy to prove that there is no
boundary effect when one applies periodic boundary conditions, e.g., an
ideal gas confined on a torus.

\section*{V. Exact solutions for photon gases in three-dimensional cavities}

In this section we calculate the exact solution for black-body radiation in
confined space. From the electromagnetic field theory, it is found that the
allowed values of ${\bf k}$ (the momentum ${\bf p}=\hbar {\bf k}$ ) in a
rectangular cavity of sides $L_{x}$, $L_{y}$ and $L_{z}$ are

\begin{equation}
k_{x}=\frac{\pi }{L_{x}}n_{x},~k_{y}=\frac{\pi }{L_{y}}n_{y},~k_{z}=\frac{%
\pi }{L_{z}}n_{z},~~~~n_{x},n_{y},n_{z}=0,1,2\cdots ,
\end{equation}%
and the energy of a photon is

\begin{equation}
E(n_{x},n_{y},n_{z})=cp=\pi \hbar c\sqrt{\frac{n_{x}^{2}}{L_{x}^{2}}+\frac{%
n_{y}^{2}}{L_{y}^{2}}+\frac{n_{z}^{2}}{L_{z}^{2}}},
\end{equation}%
where $c$ is the velocity of light. Note that, $(n_{x},n_{y},n_{z})\neq
(1,0,0)$, $(0,1,0)$, $(0,0,1)$, or $(0,0,0)$, because if any two of the
integers $n_{x},n_{y},n_{z}$ are zero, then all of the components of the
electromagnetic field will be zero. Since the number of photons is
indefinite, then the chemical potential of the system is identically zero,
we have%
\begin{equation}
\ln \Xi =2\left[ \sum\limits_{n_{x}=1}^{\infty
}\sum\limits_{n_{y}=1}^{\infty }\sum\limits_{n_{z}=1}^{\infty
}A(n_{x},n_{y},n_{z})+A(1,1,0)+A(1,0,1)+A(0,1,1)\right] ,  \label{e6.2}
\end{equation}%
where

\begin{equation}
A(n_{x},n_{y},n_{z})=-\ln \left\{ 1-\exp \left[ -\beta E(n_{x},n_{y},n_{z})%
\right] \right\}
\end{equation}%
and the factor $2$ comes from the two possible polarizations. The summation
in Eq. (\ref{e6.2}) can be performed by using the Euler-MacLaurin formula:

\begin{eqnarray}
\ln \Xi &=&2\left[ \frac{V}{\lambda ^{3}}\frac{\Gamma (4)}{\Gamma (5/2)}%
\zeta (4)-\frac{1}{4}\frac{S}{\lambda ^{2}}\frac{\Gamma (3)}{\Gamma (2)}%
\zeta (3)+\frac{1}{16}\frac{L}{\lambda }\frac{\Gamma (2)}{\Gamma (3/2)}\zeta
(2)+A(1,1,0)+A(1,0,1)+A(0,1,1)\right]  \nonumber \\
&=&V\frac{\pi ^{2}}{45\hbar ^{3}c^{3}}(kT)^{3}-S\frac{1}{4\pi \hbar ^{2}c^{2}%
}\zeta (3)(kT)^{2}+L\frac{\pi }{48\hbar c}kT+2\left[
A(1,1,0)+A(1,0,1)+A(0,1,1)\right] ,  \label{e6.3}
\end{eqnarray}%
where, as defined in Sec. III, $S$ is the area of the surface and $L$ the
total length of the sides of the cavity, $\zeta (n)$ is the Riemann zeta
function, $\lambda =2\sqrt{\pi }\hbar c/(kT)$ is the mean wavelength of the
photon.

For the total energy in the cavity, from Eq. (\ref{e6.3}) we obtain

\begin{eqnarray}
U &=&-\frac{\partial }{\partial \beta }\ln \Xi  \nonumber \\
&=&V\frac{\pi ^{2}}{15\hbar ^{3}c^{3}}\left( kT\right) ^{4}-S\frac{\zeta (3)%
}{2\pi \hbar ^{2}c^{2}}\left( kT\right) ^{3}+L\frac{\pi }{48\hbar c}\left(
kT\right) ^{2}+2\left[ B(1,1,0)+B(1,0,1)+B(0,1,1)\right] ,  \label{e6.6}
\end{eqnarray}%
where

\begin{equation}
B(n_{x},n_{y},n_{z})=\frac{E(n_{x},n_{y},n_{z})}{\exp [\beta
E(n_{x},n_{y},n_{z})]-1}.
\end{equation}
It is easy to see that the first term of Eq. (\ref{e6.6}) corresponds to the
Stefan law of black-body radiation: $U\propto T^{4}$ \cite{Huang}. It
follows that the specific heat is

\begin{equation}
C_{V}=k\left\{ V\frac{4\pi ^{2}}{15\hbar ^{3}c^{3}}\left( kT\right) ^{3}-S%
\frac{3\zeta (3)}{2\pi \hbar ^{2}c^{2}}\left( kT\right) ^{2}+L\frac{\pi }{%
24\hbar c}\left( kT\right) +2\left[ D(1,1,0)+D(1,0,1)+D(0,1,1)\right]
\right\} ,  \label{e6.8}
\end{equation}%
where
\begin{equation}
D(n_{x},n_{y},n_{z})=\frac{\left[ \beta E(n_{x},n_{y},n_{z})\right] ^{2}\exp
[\beta E(n_{x},n_{y},n_{z})]}{\{\exp [\beta E(n_{x},n_{y},n_{z})]-1\}^{2}}.
\end{equation}
The contribution from the first term, which is proportional to the volume of
the system, is just the result obtained under the hypothesis that $%
V\rightarrow \infty $ (the thermodynamic limit). The last three terms, which
correspond to the contributions from certain states, are so small that can
be neglected.

Furthermore, from the result obtained above we can learn that the density of
states of a photon gas in confined space can be approximately written as

\begin{equation}
\rho (\omega )d\omega =\left( V\frac{1}{\pi ^{2}c^{3}}\omega ^{2}-\frac{1}{4}%
S\frac{1}{\pi c^{2}}\omega \right) d\omega ,
\end{equation}%
where the second term describes the boundary effect. In other words, one can
replace the summation over states $\sum_{s}$ by the integral $\int \rho
(\omega )d\omega $.

\section*{VI. Phonon gases in confined space}

In this section, we discuss the boundary effect on phonon gases. Solids are
crystal lattices of atoms, each atom behaving as an coupled harmonic
oscillator in the harmonic approximation. Phonons correspond to the normal
modes of the system and behave as independent quantum oscillators. Assume
that the system is a rectangular solid of dimensions $L_{x}$, $L_{y}$, $%
L_{z} $. The wavevector of the phonon ${\bf k}$ is restricted by fixed-end
boundary conditions (not periodic boundary conditions because the system we
considered is finite) to the values

\begin{equation}
\left\{ \matrix{ \displaystyle k_{x}=\frac{\pi }{L_{x}},\frac{2\pi
}{L_{x}},\cdots ,\frac{ N_{x}\pi }{L_{x}}, \cr \displaystyle k_{y}=\frac{\pi
}{L_{y}},\frac{2\pi }{L_{y}},\cdots ,\frac{ N_{y}\pi }{L_{y}}, \cr
\displaystyle k_{z}=\frac{\pi }{L_{z}},\frac{2\pi }{L_{z}},\cdots ,\frac{
N_{z}\pi }{L_{z}}, }\right.
\end{equation}%
where $N_{i}$, $i=x,y,z$, is the number of atoms along $i$-axis, and the
total number of atoms $N=N_{x}N_{y}N_{z}$. Strictly speaking, there are only
$N_{i}-1$, not $N_{i}$, allowed values of $k_{i}$, because the solution for $%
N_{i}\pi /L_{i}$ permits no motion of any atom \cite{Kittel}; however, such
an influence is so small that can be neglected safely.

The grand potential is

\begin{eqnarray}
\ln \Xi &=&-\beta \left( \phi _{0}+\sum_{i=1}^{3N}\frac{1}{2}\hbar \omega
_{i}\right) -\sum_{i=1}^{3N}\ln \left( 1-e^{-\beta \hbar \omega _{i}}\right)
\nonumber \\
&=&-\beta \left[ \phi _{0}+\sum_{i=1}^{N}\frac{1}{2}\hbar (c_{l}+2c_{t})k_{i}%
\right] -\sum_{i=1}^{N}\ln \left( 1-e^{-\beta \hbar c_{l}k_{i}}\right)
-2\sum_{i=1}^{N}\ln \left( 1-e^{-\beta \hbar c_{t}k_{i}}\right) ,
\end{eqnarray}%
where
\begin{equation}
\omega _{l}=c_{l}\left\vert k\right\vert ,\text{ \ }\omega
_{t}=c_{t}\left\vert k\right\vert .
\end{equation}
There are three polarizations for each value of $k$: two of these are
transverse and one longitudinal. $c_{t}$ is the velocity of transverse waves
and $c_{l}$ the velocity of longitudinal waves. The summations in $\ln \Xi $
can be converted to a set of integrals by using the Euler-MacLaurin formula
Eq. (\ref{e2.6}). Neglecting higher orders, we have

\begin{eqnarray}
\ln \Xi  &=&-\frac{\phi _{0}}{kT}-N\left( \frac{\theta _{p}}{\theta _{0}}%
\right) ^{3}\left[ \frac{9}{8}\frac{\theta _{p}}{T}+3\ln \left( 1-e^{-\frac{%
\theta _{p}}{T}}\right) -D_{3}\left( \frac{\theta _{p}}{T}\right) \right]
\nonumber \\
&&+\frac{S}{a^{2}}\frac{\pi }{16}\left[ \frac{1}{3}\left( \theta
_{l}+2\theta _{t}\right) \frac{1}{T}+\ln \left( 1-e^{-\frac{\theta _{l}}{T}%
}\right) -\frac{1}{2}D_{2}\left( \frac{\theta _{l}}{T}\right) +2\ln \left(
1-e^{-\frac{\theta _{t}}{T}}\right) -D_{2}\left( \frac{\theta _{t}}{T}%
\right) \right] ,  \label{e9.2}
\end{eqnarray}%
where
\begin{equation}
\theta _{p}=\frac{\hbar \omega _{p}}{k},\text{\ }\theta _{l}=\frac{\pi \hbar
c_{l}}{ka},\text{ }\theta _{t}=\frac{\pi \hbar c_{t}}{ka},
\end{equation}%
and $a$ is the lattice constant. Like the Debye frequency in the Debye
approximation we introduce the cutoff frequency $\omega _{p}$, which will
return to the Debye frequency $\omega _{D}$ so that $\theta _{p}$ will
return to the Debye temperature $\theta _{D}$ when one neglects the
contribution from the boundary effect. As an approximation, we can take $%
\omega _{p}\approx $ $\omega _{D}$. Here we define the function $D_{\sigma
}(x)$ by

\begin{equation}
D_{\nu }(x)=\frac{\nu }{x^{\nu }}\int\nolimits_{0}^{x}d\xi \frac{\xi ^{\nu }%
}{e^{\xi }-1};
\end{equation}%
the case of $\nu =3$ gives the Debye function.

From Eq. (\ref{e9.2}), we can calculate the internal energy

\begin{equation}
U=\phi _{0}+NkT\left( \frac{\theta _{p}}{\theta _{0}}\right) ^{3}\left[
\frac{9}{8}\frac{\theta _{p}}{T}+3D_{3}\left( \frac{\theta _{p}}{T}\right) %
\right] -\frac{S}{a^{2}}kT\frac{\pi }{16}\left[ \frac{1}{3}\left( \frac{%
\theta _{l}}{T}+2\frac{\theta _{t}}{T}\right) +D_{2}\left( \frac{\theta _{l}%
}{T}\right) +2D_{2}\left( \frac{\theta _{t}}{T}\right) \right] ,
\end{equation}%
and the specific heat

\begin{equation}
C_{V}=3Nk\left( \frac{\theta _{p}}{\theta _{0}}\right) ^{3}\left[
4D_{3}\left( \frac{\theta _{p}}{T}\right) -3\frac{\theta _{p}/T}{e^{\theta
_{p}/T}-1}\right] -\frac{S}{a^{2}}k\frac{\pi }{16}\left[ 3D_{2}\left( \frac{%
\theta _{l}}{T}\right) -2\frac{\theta _{l}/T}{e^{\theta _{l}/T}-1}%
+6D_{2}\left( \frac{\theta _{t}}{T}\right) -4\frac{\theta _{t}/T}{e^{\theta
_{t}/T}-1}\right] .
\end{equation}

In the high temperature limit ($T\gg \theta _{p},\theta _{l},\theta _{t}$ or
$x\ll 1$), the internal energy and the specific heat are

\begin{equation}
U=\phi _{0}+3NkT\left( \frac{\theta _{p}}{\theta _{0}}\right) ^{3}\left[ 1+%
\frac{1}{20}\left( \frac{\theta _{p}}{T}\right) ^{2}\right] -\frac{S}{a^{2}}%
kT\frac{3\pi }{16}\left[ 1+\frac{1}{72}\frac{\theta _{l}^{2}+2\theta _{t}^{2}%
}{T^{2}}\right] ,
\end{equation}%
and

\begin{equation}
C_{V}=3Nk\left( \frac{\theta _{p}}{\theta _{0}}\right) ^{3}\left[ 1-\frac{1}{%
20}\left( \frac{\theta _{p}}{T}\right) ^{2}\right] -\frac{S}{a^{2}}k\frac{%
3\pi }{16}\left[ 1-\frac{1}{72}\frac{\theta _{l}^{2}+2\theta _{t}^{2}}{T^{2}}%
\right] ,  \label{e9.5}
\end{equation}%
where
\begin{equation}
\theta _{0}^{3}=18\pi ^{2}\frac{N}{V}\left( \frac{\hbar }{k}\right)
^{3}\left( \frac{1}{c_{l}^{3}}+\frac{2}{c_{t}^{3}}\right) ^{-1}
\end{equation}%
and $\theta _{0}$ equals the Debye temperature.

Correspondingly, in the low temperature region ($T\ll \theta _{p},\theta
_{l},\theta _{t}$ or $x\gg 1$) the internal energy and the specific heat
become

\begin{equation}
U=\phi _{0}+Nk\left[ \frac{9}{8}\left( \frac{\theta _{p}}{\theta _{0}}%
\right) ^{3}\theta _{p}+\frac{3\pi ^{4}}{5}\frac{1}{\theta _{0}^{3}}T^{4}%
\right] -\frac{S}{a^{2}}k\left[ \frac{\pi }{48}\left( \theta _{l}+2\theta
_{t}\right) +\frac{\pi }{4}\zeta (3)\left( \frac{1}{\theta _{l}^{2}}+\frac{2%
}{\theta _{t}^{2}}\right) T^{3}\right] ,
\end{equation}%
and

\begin{equation}
C_{V}=Nk\frac{12\pi ^{4}}{5}\frac{T^{3}}{\theta _{0}^{3}}-\frac{S}{a^{2}}k%
\frac{3\pi }{4}\zeta (3)\left( \frac{1}{\theta _{l}^{2}}+\frac{2}{\theta
_{t}^{2}}\right) T^{2}.  \label{e9.8}
\end{equation}

Each expression of the thermodynamic quantities has been divided into two
parts: the first part, though it is also influenced by the boundary, is
almost the result given by Debye; the second part, which is proportional to
the area of the surface, describes the boundary effect.

Moreover, similar to the case of ideal gases in annular containers, we can
easily prove that there is no contribution coming from the term which is
proportional to the area of the surface of the system if we apply periodic
boundary conditions.

\section*{VII. Geometry effects in realistic systems}

In this section, we shall briefly discuss the geometry effects in realistic
systems. Such effects are usually negligible in the thermodynamic limit ($%
V\longrightarrow \infty $). The subject which will be of most interest to us
will be that of boundary effects on the thermodynamic property of the system
in which the thermal wavelength of the particle is comparable to the system
size, e.g., mesoscopic systems or optical microcavities.

{\it 1) Small size systems:} In statistical physics, one often uses the
thermodynamic limit ($V\longrightarrow \infty $) as a convenient
mathematical device in macroscopic limit. The system approaches the
macroscopic limit once its size is much larger than the thermal wavelength.
The size of a mesoscopic system is between microscopic and macroscopic
scale. In mesoscopic systems, of course, many of rules in macroscopic
physics, such as the thermodynamic limit (which neglects the boundary
effect), may not hold. Many novel phenomena exist that are intrinsic to
mesoscopic systems, e.g., size effects \cite{Imry}.

In realistic systems, the magnitude of boundary effects is determined by the
ratio $\lambda /L$, where $\lambda $ is the thermal wavelength and $L$
denotes the linear size of the system. The thermal wavelength depends on two
factors: the temperature of the system $T$ and the mass of the particle. For
electrons, from room temperature to $10K$ the thermal wavelength is in the
range of about $4nm$ to $24nm$. This means that in a nanosystem the boundary
effect becomes important and can not be neglected. Furthermore, in
ultra-low-temperature physics, obviously, the boundary effect will become
remarkable. Up to now, the lowest temperature can be obtained in experiments
is of the order of $nK$ \cite{Leanhardt}. In this ultra-low temperature
scale, the wavelength of a electron is of the order of $1mm$, the wavelength
of a hydrogen atom is of the order of $0.1mm$; in other words, in such a
case, the boundary effect can be measured in macroscale.

In the case of ideal gases confined in annular containers or crystal
lattices with periodic boundary conditions, however, we have shown that
every system, very large or very small, will not show any boundary effects,
that is to say even in the small systems like torus $C_{360}$ and $C_{240}$
molecules, if we can still treat such systems by the statistical method,
there are no boundary effects so long as the systems have the toroidal
structure. Recently, there are many studies on nanorings, especially the
metal nanorings \cite{Bagci}. It might be expected that the method and the
result of ideal gases in annular containers can be used to treat the
electron gases in metal nanorings. Moreover, metal wires having diameters in
the range of nanometer are very important for nanoelectronics and other
nanodevice applications, and, recently, ultra-thin metal nanowires have
aroused growing interest in condensed matter physics \cite{Ciraci}. The
analysis presented above applies equally to such systems.

Of special interest is the electron gas on a carbon nanotube. It is known
that the electronic structure of a carbon nanotube can be either metallic or
semiconducting, depending on its diameter and chirality \cite{Saito}. For
the metallic case we can treat the electrons on a nanotube as a free
electron gas on the lateral surface of a cylinder. In such a system, along
the perimeter of the transverse cross-section, even the size is very small
so that the energy spectrum of the electrons is discrete, there is no
boundary effect (see Sec. IV); otherwise, along the nanotube axis the
boundary effect correction has to be reckoned in when the length of the
nanotube is very short. With the help of the method provided in Sec. III we
can calculate the thermodynamic properties of such systems directly, e.g.
the specific heat in Eq. (\ref{e2.18}).

{\it 2) Long and narrow systems:} Simple analysis reveals that the effect of
boundary on the observable is proportional to $L/\sqrt{S}$, where $S$ is the
area and $L$ the perimeter of the system. The magnitude of the factor $L/%
\sqrt{S}$ is determined by the geometry of the system. If the shape of the
system is long and narrow, in principle, the factor $L/\sqrt{S}$ can take on
an arbitrarily large value and hence the boundary effect will become
significant even in macroscale.

{\it 3) Black-body radiation in small cavities:} From visible light to
infrared, the range of the wavelength is from $0.4$ to about $10^{2}$
micron. This means that if the size of a system is of the order of micron,
the boundary effect must be reckoned in. In Sec. V we have calculated the
exact solution for black-body radiation in a three-dimensional rectangular
cavity, from which we can estimate the influence of the boundary effect on
black-body radiation.

From Eq. (\ref{e6.8}) we can estimate the influence of the boundary effect
on the specific heat of a photon gas confined in a small cavity by comparing
the contribution from the surface term (the second term of Eq. (\ref{e6.8}))
with the result which is obtained based on the thermodynamic limit (the
first term of Eq. (\ref{e6.8})). For this purpose, we introduce a ratio $%
\eta =C_{V}^{S}/C_{V}^{V}$, where $C_{V}^{S}$ is defined as the second term
and $C_{V}^{V}$ the first term of Eq. (\ref{e6.8}). For the linear size of
the system $L=1mm$ and $0.1mm$, when the temperature $T=300K$, the ratios $%
\eta $ are about $1\%$ and $10\%$; when $T=100K$, the ratios $\eta $ are
about $3\%$ and $30\%$; moreover, when $T=10K$, $L=1cm$, the ratio $\eta
\sim 3\%$. This result shows that, relative to the case of particles with
non-zero rest masses (whose wavelengths are relatively short), the boundary
effect is significant in photon systems.

{\it 4) Systems consisting of small grains:} In Sec. VI we have discussed
the boundary correction to the Debye theory of the specific heat of
crystals. From Eq. (\ref{e9.8}) we can learn that at low temperature the
boundary effect will become significant since the contribution from the
boundary term is proportional to $T^{2}$, whereas the term corresponding to
the Debye's result is proportional to $T^{3}$. Therefore, the contribution
from the boundary effect will become more and more important with a drop in
temperature. Of course, the smaller the system size, the stronger is the
influence of the boundary. Consider a macroscopic system consisting of small
grains. Such a granular material can be obtained, for example, by grinding a
piece of crystal to a fine powder. In such a system the volume is almost the
same as the volume before grinding, whereas the total area of surface, which
equals the sum of the area of all grains in the system, becomes much larger.
We can rewrite the expression of the specific heat in the following form for
clarity:

\begin{equation}
C_{V}=\frac{V}{a^{3}}A(T)-\frac{S}{a^{2}}B(T),
\end{equation}%
where the area of surface $S\sim V^{2/3}$, $A(T)$ and $B(T)$ are certain
functions of the temperature. If we divide the volume $V$ up into $n$ cells
of volume $V/n$, the total volume of the system is roughly speaking still $V$%
; however, the total area of surface becomes $n\left( V/n\right) ^{2/3}=\sqrt%
[3]{n}V^{2/3}$. This means that the boundary effect will become important in
granular materials. It can be found that before grinding the correction to
the standard result (the term which is proportional to $V$) is on the order $%
a/\sqrt[3]{V}$, here we have not reckoned in the influence of temperature
though it is of equal importance, whereas after grinding such a correction
increases to the order $\sqrt[3]{n}\left( a/\sqrt[3]{V}\right) $.

{\it 5) Geometry effects vs. fluctuations:} The geometry effect may be
visible for small size systems. In such systems, however, the fluctuations
also may not be neglected since the total number of particles $N$ is not
macroscopic. A natural question that may be asked is: Can one distinguish
geometry effects from the fluctuation background? Based on the results
calculated above, we can estimate in what cases the geometry effects will
become apparent. Next we consider some special cases as examples.

The influence of fluctuations is on the order $\sqrt{N}/N$. In metals, take
copper as an example, one $cm^{3}$ contains about $10^{23}$ free electrons.
The ratio $\mu _{0}^{boundary}/\mu _{0}^{free}$ reflects the influence of
the boundary on the Fermi energy. When the linear size of the system is $5nm$%
, the ratio $\mu _{0}^{boundary}/\mu _{0}^{free}\sim 3.3\%$; however, in
this case the fluctuation is only about $0.9\%$; when the size of the system
is $30nm$, the ratio $\mu _{0}^{boundary}/\mu _{0}^{free}\sim 0.6\%$, but
the fluctuation is about $0.06\%$. This means that the magnitude of the
fluctuation, though also can not be neglected, is sufficiently smaller than
that of the boundary effect so that one can distinguish the boundary effect
in nanoscale metal systems.

For photon gases with linear size $L=1mm$ (macroscale) and temperature $%
T=300K$ (room temperature), the ratio $\eta $ ($=C_{V}^{S}/C_{V}^{V}$)$\ $is
about $1\%$, but, in this case, the fluctuation is negligibly small; for the
case of $L=1cm$ and $T=10K$ the ratio $\eta \sim 3\%$, but the fluctuation
is only about $0.1\%$. That is to say, the boundary effect can be measured
in macroscopic photon systems.

{\it 6) Geometry effects vs. particle-particle interactions:} In most
realistic systems, there will be interactions between the particles. Next,
we compare the contribution from the geometry effects with the influence of
the interactions among the particles.

(a) Bose case: For dilute Bose gases, the weak interactions between
particles and their low density allow for an accurate theoretical
description of the effect of particle-particle interactions based on the
binary collision method. When $T>T_{c}$, where $T_{c}$ is the critical
temperature, the equation of state can be expressed as \cite{Lee}%
\begin{equation}
\frac{P}{kT}=\frac{1}{\lambda ^{3}}g_{5/2}(z)-\frac{2a}{\lambda ^{4}}%
g_{3/2}^{2}(z)+\cdots ,
\end{equation}%
where $a$ is the scattering length; the influence of the particle-particle
interactions is described in the second term. For geometry effects, Eq. (\ref%
{e3.18}) gives the equation of state for an ideal Bose gas in confined space:

\begin{equation}
\frac{P}{kT}=\frac{1}{\lambda ^{3}}g_{5/2}(z)-\frac{1}{4}\frac{S}{V\lambda
^{2}}g_{2}(z)+\cdots .
\end{equation}%
From this result we can estimate the contributions from the geometry effects
and the interparticle interactions: The leading contribution from the
boundary is $-(1/4)S/(V\lambda ^{2})g_{2}(z)$; the leading contribution from
the interparticle interaction is $-(2a/\lambda ^{4})g_{3/2}^{2}(z)$. By
comparing these two contributions, we can see that there exists a regime
where the geometry effect dominates in an imperfect gas. Introduce the ratio
between these two contributions

\begin{equation}
\eta _{b}=\frac{1}{8}\frac{S\lambda ^{2}}{Va}\frac{g_{2}(z)}{g_{3/2}^{2}(z)},
\end{equation}%
where $a$ denotes the magnitude of the scattering length. The regime that
the geometry effect becomes dominate can be determined by the condition $%
\eta _{b}>1$.

Using this result, we next compare the influence of the geometry effects
with the influence of the interparticle interactions on Bose-Einstein
condensation. We will show that the geometry effects, compared with the
interparticle interactions, are important in some recent experiments on
Bose-Einstein condensation in dilute atomic gases.

In the recent experiments on Bose-Einstein condensation, the shape of the
condensate is either cigar-shaped, with a diameter about $15\mu m$ and
length $300\mu m\sim 5mm$, or approximately round with a diameter of $10\sim
50\mu m$ \cite{Anglin,Kleppner}. As an example, the Bose gas we considered
here is near the critical point where the fugacity $z\sim 1$ and the
temperature $T\sim T_{c}$, and the shape of the corresponding condensate is
cigar-like (or thread-like), in other words, the shape of the system is long
and narrow. In such a system, the total area of the surface is approximately
equal to the area of the lateral surface, so we have%
\begin{equation}
d_{\max }\sim 0.12\frac{\lambda ^{2}}{a},
\end{equation}%
where $d_{\max }$ is the maximum size of the transverse cross-section of the
system in which the geometry effect becomes dominate. The parameters in some
recent experiments have the following orders of magnitude \cite{Kleppner}:
For $^{1}H$, the scattering length $a\sim 0.065nm$, the thermal wavelength $%
\lambda \sim 0.25\mu m$, so we have $d_{\max }\sim 0.11mm$; for $^{7}Li$, $%
a\sim -1.5nm$, $\lambda \sim 1.2\mu m$, so $d_{\max }\sim 0.12mm$; for $%
^{23}Na$, $a\sim 2.8nm$, $\lambda \sim 0.26\mu m$, so $d_{\max }\sim
2.9\times 10^{-3}mm$; for $^{87}Rb$, $a\sim 5.4nm$, $\lambda \sim 0.45\mu m$%
, so $d_{\max }\sim 4.3\times 10^{-3}mm$. From this result we can see that,
compared with the interparticle interactions, the geometry effects can not
be ignored in the experiments on Bose-Einstein condensation of $^{1}H$ and $%
^{7}Li$. For example, in the experiment on Bose-Einstein condensation of
atomic hydrogen the condensate is $15\mu m$ in diameter and $5mm$ in length
\cite{Kleppner}, so the ratio $\eta _{b}\sim 7.5$, i.e., the geometry
effects are more important than the interparticle interactions in the
experiment on Bose-Einstein condensation of atomic hydrogen.

Strictly speaking, in the recent experiments on Bose-Einstein condensation
the atomic gases are trapped in magnetic traps rather than boxes with hard
walls. Usually, such traps are approximately described as effective
three-dimensional harmonic wells cylindrically symmetric about the $z$-axis.
In above analysis we assume that the Bose gases are trapped in hard-wall
tubes, i.e., for simplifying the treatment, we replace the harmonic wells by
tubes, but the result, to orders of magnitude, is reliable.

(b) Fermi case: For an imperfect Fermi gas the equation of state is \cite%
{Lee}

\begin{equation}
\frac{P}{kT}=\left( 2j+1\right) \frac{1}{\lambda ^{3}}f_{5/2}(z)-2j(2j+1)%
\frac{a}{\lambda ^{4}}f_{3/2}^{2}(z)+\cdots ,  \label{e18.8}
\end{equation}%
where $j$ is the spin of the particle. Like the treatment in the Bose case,
we can also compare the geometry effects with the interparticle interactions
by comparing the second terms in Eqs. (\ref{e3.18}) and (\ref{e18.8})
directly. Next, as an example, we compare these two influences on the ground
state of a Fermi system.

The ground-state energy for an imperfect Fermi gas is given by ($j=1/2$)
\cite{Lee,Huang2}

\begin{equation}
E=\frac{3}{5}N\frac{\hbar ^{2}}{2m}\left( 3\pi ^{2}\frac{N}{V}\right)
^{2/3}+N\frac{\hbar ^{2}}{2m}2\pi a\frac{N}{V}+\cdots ,
\end{equation}%
the contribution from the interparticle interactions described in the second
term. The corresponding chemical potential is

\begin{eqnarray}
\mu _{0} &=&\frac{\hbar ^{2}}{2m}\left( 3\pi ^{2}\frac{N}{V}\right) ^{2/3}+%
\frac{\hbar ^{2}}{2m}\left( 4\pi a\frac{N}{V}\right)  \nonumber \\
&=&\mu _{0}^{free}+\mu _{0}^{interparticle},
\end{eqnarray}%
where $\mu _{0}$ is equal to the Fermi energy of the ideal Fermi gas plus
the second term which describes the contribution from the interparticle
interactions. Similarly, as expressed in Eq. (\ref{e3.38}), the Fermi energy
containing the contribution from the boundary can be written as $\mu
_{0}=\mu _{0}^{free}+\mu _{0}^{boundary}$. Introducing a ratio $\eta
_{f}=\mu _{0}^{boundary}/\mu _{0}^{interparticle}$, we can compare the
corrections due to the geometry effects with the corrections that emerge
from the interactions among the particles. Direct calculation gives

\begin{equation}
\eta _{f}=\frac{1}{16}(3\pi ^{2})^{1/3}\frac{1}{a}\frac{S}{V}\left( \frac{N}{%
V}\right) ^{-2/3}.
\end{equation}%
Taking a box of side-length $d$ as an example, we have $\eta _{f}=(3/8)(3\pi
^{2})^{1/3}l^{2}/\left( ad\right) $, where $l=\left( V/N\right) ^{1/3}$ is
the mean interparticle distance. Obviously, if $\eta _{f}>1$, compared with
the interparticle interactions, the geometry effects will dominate in the
system. $\eta _{f}>1$ gives the condition
\begin{equation}
\frac{d}{l}<1.160\frac{l}{a}.
\end{equation}%
This condition is equivalent to the requirement of low densities and small
system sizes. That is to say, in a small dilute Fermi gas, the influence of
geometry effects will exceed the influence of interparticle interactions.

Furthermore, in this paper we also consider the geometry effects on photon
gases and on phonon gases. The photon gas can be regarded as a genuine ideal
gas in statistical mechanics since the cross section of photon-photon
scattering, which is only a loop correction, is exceedingly small.
Therefore, there is no contribution from the interparticle interaction. In
the harmonic approximation, the phonon gas can also be regarded as an ideal
gas. That is to say, in photon gases and in phonon gases the contributions
from the geometry effects are the main corrections to the standard results.

\section*{VIII. Discussions and conclusions}

The key problem in statistical mechanics is to calculate the partition
function, i.e., to solve the sum over all possible states. Often, this sum
is difficult to calculate. To calculate the partition function
approximately, in the thermodynamic limit $V\rightarrow \infty $, one can
convert the sum to an integral by introducing a state density. In so doing,
however, the information about the system geometry has been lost because the
approximation that the spectrum is continuous has been taken during this
process. The information about the system geometry is embodied in the
structure of the spectrum, so if we want to involve the geometry information
in the thermodynamic quantities, we have to solve the sum over the states
directly. We have developed two methods to solve the sum: One is an
approximate method which can be used to deal with the system with an
irregular boundary \cite{Ours}. Another method considered in the present
paper is based on the Euler-MacLaurin formula; sometimes this method leads
to exact solutions. However, the prerequisite for applying the
Euler-MacLaurin formula is that the spectrum is already known.

The method for performing the summation by the Euler-MacLaurin formula is
easy to apply to the cases of ideal gases in external potentials and
nonideal gases if the spectrum of the system is already known. Fortunately,
the spectrum can be obtained by using many systematic methods exactly or
approximately. Such cases we will discuss in detail elsewhere \cite{Ours2}.

Intrinsically the boundary effect arises from the interaction between the
particles and the boundary. The influence of this interaction can be
embodied in boundary conditions. The fix-end boundary condition, like the
case of a particle in an infinite depth potential, reflects the interaction
between the particles and the wall of the box. However, if there is no
boundary, of course, there is no interaction between the particles and the
boundary; in other words, there is no boundary effect. In such a case we may
apply periodic boundary conditions, and then the boundary terms will vanish
in the exact solution.

From the results given above, we can see that the grand potential of a
system in confined space is less than that in free space since the sign of
the leading geometry term, e.g., the second term in Eq. (\ref{e3.8}), is
negative. It means that the existence of a boundary tends to reduce the
number of states of the system, in other words, to reduce the volume of the
phase space --- the set of all possible states. This is just because, for
ideal gases, in free space the spectrum is continuous while in confined
space the spectrum gets discrete. Thus the number of modes in confined space
is less than that in free space.

The boundary term which is proportional to the area of the surface is always
negative (the other terms corresponding to geometry effects are usually
negligible). A natural question that may be asked is: What will happen when
the contribution from the boundary effect become large enough, e.g., in the
case of the size of the system is very small or the temperature is very low,
so that the grand potential is less than zero? The answer is that in such a
case the statistical method is no longer valid because the grand potential
is the logarithm of the grand partition function and a negative grand
potential means that the number of the states is less than one.

In conclusion, by using the Euler-MacLaurin formula, the geometry effects on
the statistical mechanics of various systems, such as ideal Bose and Fermi
gases, photon gases, phonon gases, are discussed. From the exact solutions
we can see that the grand potential can be expressed as a sum of two parts:
one is just the result obtained in the thermodynamic limit $V\rightarrow
\infty $; another is the contribution from the geometry effect. The latter
is consisted of a series of terms: the first term, by comparing with the
result given in Ref. \cite{Ours}, reflects the influence of the shape of the
boundary; the second term, which is proportional to the Euler-Poincar\'{e}
characteristic number, reflects the influence of the topological property of
the system; the rest terms are always negligible and the number of such
terms is finite when the problem is exactly solvable, or else it may be
infinite. Moreover, we discuss the validity of the approximate method for
calculating the effects of boundary and connectivity, presented in Ref. \cite%
{Ours}, which is based on the mathematical work given by Kac \cite{Kac} and
can be used to deal with the ideal gas system with an irregular boundary in
two dimensions. We also analyze the possible geometry effects in realistic
systems. It is hoped that the geometry effects can be encountered in the
regime of mesoscopic scale. For photon gases, the geometry effects may be
observed even in macroscale systems. In imperfect gases, we compare the
geometry effects with the interparticle interactions and point out that
there exists a regime where the geometry effects dominate in interacting
systems.

\vskip0.5cm

We are very indebted to Dr. G. Zeitrauman for his encouragement. This work
is supported in part by LiuHui fund.

\section*{Appendix A: Thermodynamic quantities in three-dimensional boxes}

Following general procedures, we can obtain the thermodynamic quantities for
an ideal gas in a three-dimensional box although the geometry terms in Eq. (%
\ref{e3.18}) make the derivation get tedious.

The equation of state:%
\begin{eqnarray*}
\frac{PV}{kT} &=&\ln \Xi =\frac{V}{\lambda ^{3}}h_{5/2}(z)-\frac{1}{4}\frac{S%
}{\lambda ^{2}}h_{2}(z)+\frac{1}{16}\frac{L}{\lambda }h_{3/2}(z)-\frac{1}{8}%
h_{1}(z), \\
N &=&z\frac{\partial }{\partial z}\ln \Xi =\frac{V}{\lambda ^{3}}h_{3/2}(z)-%
\frac{1}{4}\frac{S}{\lambda ^{2}}h_{1}(z)+\frac{1}{16}\frac{L}{\lambda }%
h_{1/2}(z)-\frac{1}{8}h_{0}(z).
\end{eqnarray*}%
Internal energy:

\[
\frac{U}{NkT}=\frac{3}{2}\frac{h_{5/2}(z)}{h_{3/2}(z)}\left( 1+\Delta
\right) +\frac{1}{N}\left[ -\frac{1}{4}\frac{S}{\lambda ^{2}}h_{2}(z)+\frac{1%
}{32}\frac{L}{\lambda }h_{3/2}(z)\right] ,
\]%
where

\[
\Delta =\frac{1}{N}\left[ \frac{1}{4}\frac{S}{\lambda ^{2}}h_{1}(z)-\frac{1}{%
16}\frac{L}{\lambda }h_{1/2}(z)+\frac{1}{8}h_{0}(z)\right] .
\]%
Specific heat:

\begin{eqnarray*}
\frac{C_{V}}{Nk} &=&\left[ \frac{15}{4}\frac{h_{5/2}(z)}{h_{3/2}(z)}-\frac{9%
}{4}\frac{h_{3/2}(z)-\delta _{1}}{h_{1/2}(z)-\delta _{2}}\right] \left(
1+\Delta \right) \\
&&-\frac{1}{2}\frac{1}{N}\frac{S}{\lambda ^{2}}\left[ h_{2}(z)-\frac{3}{4}%
h_{1}(z)\frac{h_{3/2}(z)-\delta _{1}}{h_{1/2}(z)-\delta _{2}}\right] +\frac{3%
}{64}\frac{1}{N}\frac{L}{\lambda }\left[ h_{3/2}(z)-h_{1/2}(z)\frac{%
h_{3/2}(z)-\delta _{1}}{h_{1/2}(z)-\delta _{2}}\right] ,
\end{eqnarray*}%
\newline
where%
\[
\delta _{1}=\frac{1}{6}\frac{S\lambda }{V}h_{1}(z)-\frac{1}{48}\frac{%
L\lambda ^{2}}{V}h_{1/2}(z),
\]

\[
\delta _{2}=\frac{1}{4}\frac{S\lambda }{V}h_{0}(z)-\frac{1}{16}\frac{%
L\lambda ^{2}}{V}h_{-1/2}(z)+\frac{1}{8}\frac{\lambda ^{3}}{V}h_{-1}(z).
\]

We also have

\[
\frac{\partial z}{\partial T}=-\frac{3}{2}\frac{z}{T}\frac{h_{3/2}(z)-\delta
_{1}}{h_{1/2}(z)-\delta _{2}}.
\]

\end{document}